\documentclass[twocolumn,tighten,twocolappendix]{aastex63}
\pdfoutput=1 
\usepackage[T1]{fontenc}
\usepackage{ae,aecompl}
\usepackage[utf8]{inputenc}
\usepackage[english]{babel}
\usepackage{amsmath,amstext}
\usepackage{apjfonts}
\usepackage[figure,figure*]{hypcap}
\usepackage[hang]{footmisc}
\usepackage{ulem}
\usepackage[flushleft]{threeparttable}

\lefthead{HACC Collaboration}
\righthead{The Farpoint Simulation}

\begin{document}

\title{Farpoint: A High-Resolution Cosmology Simulation at the Gigaparsec Scale}

\collaboration{10}{The HACC Collaboration}
\noaffiliation

\author{Nicholas Frontiere}
\affiliation{Computational Science Division,  Argonne National Laboratory, Lemont, IL 60439}

\author{Katrin~Heitmann}
\affiliation{High Energy Physics Division,  Argonne National Laboratory, Lemont, IL 60439}

\author{Esteban Rangel}
\affiliation{Argonne Leadership Computing Facility,  Argonne National Laboratory, Lemont, IL 60439}

\author{Patricia Larsen}
\affiliation{Computational Science Division,  Argonne National Laboratory, Lemont, IL 60439}

\author{Adrian Pope}
\affiliation{Computational Science Division,  Argonne National Laboratory, Lemont, IL 60439}

\author{Imran Sultan}
\affiliation{Department of Physics and Astronomy, Northwestern University, 2145 Sheridan Road, Evanston, IL 60208}

\author{Thomas Uram} 
\affiliation{Argonne Leadership Computing Facility,  Argonne National Laboratory, Lemont, IL 60439}

\author{Salman~Habib}
\affiliation{Computational Science Division,  Argonne National Laboratory, Lemont, IL 60439}
\affiliation{High Energy Physics Division,  Argonne National Laboratory, Lemont, IL 60439}

\author{Silvio Rizzi}
\affiliation{Argonne Leadership Computing Facility,  Argonne National Laboratory, Lemont, IL 60439}

\author{Joe Insley}
\affiliation{Argonne Leadership Computing Facility,  Argonne National Laboratory, Lemont, IL 60439}

\begin{abstract}

In this paper we introduce the Farpoint simulation, the latest member of the Hardware/Hybrid Accelerated Cosmology Code (HACC) gravity-only simulation family.
The domain covers a volume of (1000$h^{-1}$Mpc)$^3$ and evolves close to two trillion particles, corresponding to a mass resolution of $m_p\sim 4.6\cdot 10^7 h^{-1}$M$_\odot$. 
These specifications enable comprehensive investigations of the galaxy-halo connection, capturing halos down to small masses. Further, the large volume resolves scales typical of modern surveys with good statistical coverage of high mass halos.
The simulation was carried out on the GPU-accelerated system Summit, one of the fastest supercomputers currently available. We provide specifics about the Farpoint run and present an initial set of results. The high mass resolution facilitates precise measurements of important global statistics, such as the halo concentration-mass relation and the correlation function down to small scales. Selected subsets of the simulation data products are publicly available via the HACC Simulation Data Portal.

\end{abstract}



\section{Introduction}

The current era of precision cosmology targets meticulous measurements to guide our path in exploring and understanding the physics of the dark universe. Major facilities and instruments, such as the Dark Energy Spectroscopic Instrument (DESI)~\citep{desi}, the Vera C. Rubin Observatory's Legacy Survey of Space and Time (LSST)~\citep{lsst,desc}, the Nancy Grace Roman Space Telescope ~\citep{wfirst}, the Euclid satellite~\citep{euclid} and the Spectro-Photometer for the History of the Universe, Epoch of Reionization and Ices Explorer (SPHEREx)~\citep{spherex}, promise to deliver measurements of our Universe and its constituents at unprecedented statistical accuracy. The interpretation of these observations is, however, nontrivial. In particular, the understanding of various systematic effects that can potentially contaminate the data, as well as the modeling, is a major challenge.

Cosmological simulations play a critical role in addressing this problem by providing predictions and mock realizations to calibrate and quantify error tolerances of the observations and allow for explicit exploration and testing of various error modes, both observational and modeling-related. As one representative example, the LSST Dark Energy Science Collaboration (DESC) has recently generated a sophisticated end-to-end simulation to develop LSST-like catalogs~\citep{cosmoDC2,2020arXiv201005926L} in preparation for the arrival of the extraordinary data expected from that survey. 

Gravity-only N-body simulations are still the primary means for creating comprehensive synthetic sky catalogs and will continue to be viable given the computational cost and modeling uncertainties of hydrodynamical simulations.
Prominent examples of such simulations include the Euclid Flagship simulation~\citep{potter} to model the upcoming Euclid survey, the Outer Rim simulation~\citep{heitmann19b} that has been used to build catalogs for a diverse set of surveys, and the Buzzard and MICE suite of simulations for the Dark Energy Survey (DES)~\citep{2019arXiv190102401D,2015MNRAS.453.1513C}.

In recent years, significant progress has been made in pushing the limits of gravity-only simulations to ever larger volumes and improved mass resolution. However, mainly due to limitations on available system memory resources, it is still infeasible to achieve very high mass resolution ($\sim10^8$M$_\odot$ or better) in volumes large enough to fully cover survey-sized modeling requirements ($\sim$4-5Gpc). 
It is, therefore, common practice to generate sets of simulations for the same cosmology that either have high mass resolution or cover a larger volume with sufficient resolution for detailed studies of large-scale structure evolution and galaxy-halo connections; 
examples include the Millennium set of simulations~\citep{2005Natur.435..629S,2009MNRAS.398.1150B,2012MNRAS.426.2046A}, the Outer Rim/QContinuum pair~\citep{QC,heitmann19b}, 
the Horizon Runs~\citep{kim2015,lee2020},  
and the Uchuu simulation suite~\citep{2020arXiv200714720I}.

In this paper, we introduce the Farpoint simulation -- an addition to our set of available extreme-scale simulations that explore structure formation and enable important investigations closely related to upcoming cosmological surveys. The Farpoint run was carried out with HACC, the Hardware/Hybrid Accelerated Cosmology Code~\citep{habib14} on Summit, currently one of the top ranked supercomputers in the world\footnote{\url{https://www.top500.org/lists/2019/11/}}. HACC is designed to run at scale on a diverse range of computer architectures and has been individually tuned for multi-core and heterogeneous computing systems. Accordingly, HACC is optimized to take full advantage of the GPU-acceleration on Summit to achieve high performance~\citep{QC}.  

Farpoint evolves more than 1.8 trillion particles in a (1000 $h^{-1}$Mpc)$^3$ volume, leading to a high mass resolution with a particle mass of $m_p=4.6\cdot 10^7 h^{-1}$M$_\odot$. The cosmology used for the simulation is close to the best-fit Planck cosmology~\citep{planck18} and is identical to the parameters used in the Last Journey simulation~\citep{LJ1}. Collectively, this new pair of simulations cover a wide range of masses and supply excellent statistics for high-mass objects, similar in spirit to the Millennium simulation series based on WMAP-1~\citep{wmap1} and the Outer Rim/QContinuum set of simulations based on WMAP-7~\citep{wmap7}.  

\begin{figure*}[t]
\centerline{\includegraphics[width=6.4in]{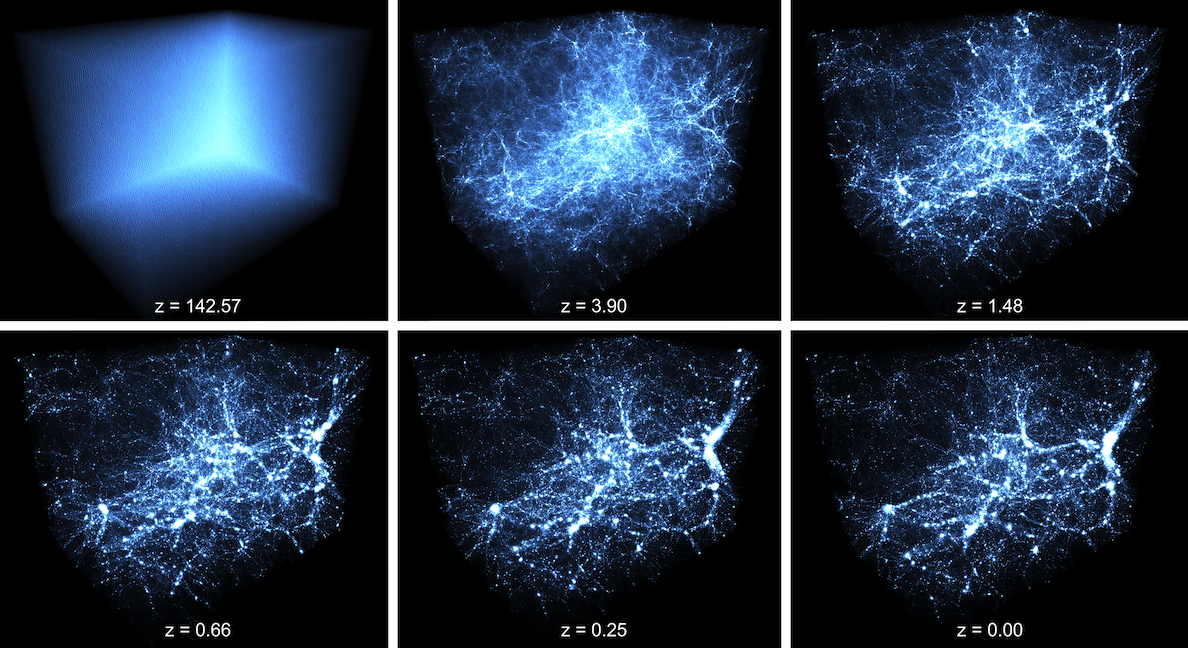}}
\caption{\label{fig:evolv} Time evolution of the Farpoint particle data captured on a single MPI rank at a number of redshifts. Since the decomposition of the simulation is not cubic (carried out on 24576 ranks, leading to a topology decomposition of $32\times32\times24$), the output is a cuboid. The visualization shows $\sim$75M particles. An animated video of this figure is available in the HTML version of the article; a 33 second clip showing the structure formation evolution of the particle data between redshifts $z=200$ and $z=0$. } 
\end{figure*}

A major scientific target for the Farpoint simulation is the creation of synthetic sky maps. In particular, the high mass resolution achieved in the Farpoint run lends itself to thorough investigations of the galaxy-halo connection (for a recent review, see,~\citealt{wechsler18}). In \cite{cosmoDC2}, the Outer Rim simulation~\citep{heitmann19a} was used to create an extragalactic catalog for cosmology studies to be carried out with the Vera Rubin Observatory. While the Outer Rim simulation covers a much larger volume (box length of 3000 $h^{-1}$Mpc) compared to the Farpoint simulation, the mass resolution was worse by a factor of $\sim 40$ and was not sufficient to accurately resolve the very faint galaxies that the Rubin Observatory will capture. Consequently, these galaxies were incorporated in that catalog via random sampling.
The Farpoint simulation will be able to render these faint galaxies more truthfully by connecting them to actual halos resolved in the simulation.

Additionally, Farpoint will facilitate the development of refined galaxy-halo connection models following the approach of \cite{2020MNRAS.495.5040H}, which in turn, will assist in improved modeling for lower-mass resolution simulations. Work in this area, such as \cite{2019MNRAS.488.3143B}, relies on high mass resolution simulations; the Farpoint run would provide an improved underpinning for such an approach. 
Moreover, the Farpoint simulation data will prove valuable when investigating fundamental questions about structure formation that require high-resolution simulations and quality statistics, such as in halo substructure studies.

The paper is organized as follows. In Section~\ref{sec:sim}, we provide details about the set-up of the simulation. A significant portion of the data analysis is carried out on the fly, as the simulation runs. We describe the results obtained from these in situ analyses in Section~\ref{sec:insitu-analysis}. Next, we showcase data products and first results obtained via post-processing pipelines in Section~\ref{sec:post-analysis}. As part of this paper, we publicly release a subset of our results. We provide specifics about those data sets and how to access them in Section~\ref{sec:data}. We conclude in Section~\ref{sec:summary} with a short summary and outlook.

\section{Simulation Specification}
\label{sec:sim}

The Farpoint simulation evolved 12,288$^3$ particles ($\sim$ 1.86 trillion) in a volume of (1000$h^{-1}$Mpc)$^3$, following a best-fit Planck 
cosmology~(\citealt{planck18}, Table 2, base-$\Lambda$CDM fit combining cosmic microwave background (CMB) spectra, with CMB lensing reconstruction and baryon acoustic oscillation (BAO) measurements). We assume spatial flatness ($\Omega_k=0$) and massless neutrinos. The six parameters specifying our $\Lambda$CDM cosmology are listed in Table~\ref{cosmoparams}, which determine a total matter contribution of $\Omega_m=0.310$.
\begin{table}[h]
\vspace{-2mm}
\begin{tabular}{ l c c }
 \textbf{Name} & \textbf{Symbol} & \textbf{Value} \\ 
 \hline
 Dark matter density & $\Omega_{\rm cdm}$ & 0.26067 \\  
 Baryon density & $\Omega_{\rm b}$ & 0.049 \\ 
 Hubble parameter & $h$ & 0.6766 \\
 Matter fluctuation amplitude & $\sigma_8$ & 0.8102 \\
 Scalar spectral index & $n_s$ & 0.9665 \\
 Dark energy EOS parameter & $w$ & -1 \\
\end{tabular}
\caption{Farpoint Cosmological Parameters}
\label{cosmoparams}
\vspace{-5mm}
\end{table}

We chose the same cosmology as for the Last Journey simulation~\citep{LJ1}, which covers a larger volume at lower mass resolution.
The specifications given above lead to a particle mass of $m_p=4.6\cdot 10^7 h^{-1}$M$_\odot$. The force resolution softening was set to $\sim 0.8h^{-1}$kpc. The initial conditions were generated using the Zel'dovich approximation~\citep{Zel70} at an initial redshift of $z_{\rm in}=200$. We employed {\sc CAMB}~\citep{camb} to create the initial matter transfer function. 

\section{In Situ Analysis}
\label{sec:insitu-analysis}

In order to fully extract the desired science from the Farpoint run, detailed investigations of cosmic structure evolution are needed. Exclusively performing the analysis in post-processing for an extreme-scale simulation poses a major challenge. Petabytes of data would need to be generated and stored, requiring a large additional compute allocation for I/O and processing the outputs. To overcome this challenge, we have developed an extensive in situ analysis toolkit within HACC called CosmoTools. The CosmoTools library provides a seamless interface to the HACC data while it resides in memory during the simulation run. Different tools can be turned on and off at specified time steps. In addition to the analyses provided by CosmoTools, we carry out several in situ tasks that are closely coupled to the time stepper, e.g., the particle light cone evaluation. These routines are driven by the HACC solver directly.  

Throughout this section we provide a brief overview of all of the different in situ analysis components carried out and list the data products generated during the simulation campaign. We also present selected measurements obtained directly from these outputs.

In order to enable comprehensive time evolution studies and the creation of synthetic sky catalogs using semi-analytic approaches, we have processed a number of analysis measurements and stored results at 101 time snapshots between $z=10$ and $z=0$, evenly spaced in $\log_{10}(a)$. This leads to the following output values in redshift:
\begin{eqnarray}
z&=&\left\{10.04, 9.81, 9.56, 9.36, 9.15, 8.76, 8.57, 8.39, 8.05,
   \right.\nonumber\\ 
&&7.89, 7.74, 7.45, 7.31, 7.04, 6.91, 6.67, 6.56, 6.34, 6.13, 
\nonumber\\ 
&&6.03, 5.84, 5.66, 5.48, 5.32, 5.24, 5.09, 4.95, 4.74, 4.61, 
\nonumber\\ 
&&4.49, 4.37, 4.26, 4.10, 4.00, 3.86, 3.76, 3.63.  3.55, 3.43, \nonumber\\ 
&&3.31, 3.21, 3.10,
3.04, 2.94, 2.85, 2.74, 2.65, 2.58, 2.48, 
\nonumber\\ 
&&2.41, 2.32, 2.25, 2.17, 2.09, 2.02, 1.95, 1.88, 1.80, 1.74, 
\nonumber\\ 
&&1.68, 1.61, 1.54, 1.49, 1.43, 1.38, 1.32, 1.26, 1.21, 1.15, 
\nonumber\\ 
&&1.11, 1.06, 1.01, 0.96, 0.91, 0.86, 0.82, 0.78, 0.74, 0.69, 
\nonumber\\ 
&&0.66, 0.62, 0.58, 0.54, 0.50, 0.47, 0.43, 0.40, 0.36, 0.33, 
\nonumber\\ 
&&0.30, 0.27, 0.24, 0.21, 0.18, 0.15, 0.13, 0.10, 0.07, 0.05, 
\nonumber\\ 
&&\left.
0.02, 0.00\right\}.
\label{redshifts}
\end{eqnarray}
We have used the same strategy for several of our other major simulations to facilitate easy comparison of results across the different runs. All of our outputs utilize lossless compression via the Blosc library.\footnote{\url{https://blosc.org/}} The compression approach leads to a reduction of the data by approximately 2x, depending on the data product. \cite{LJ1} provide a more thorough discussion about the compression factors achieved for different outputs.

\subsection{Particle Snapshots}
\label{sec:part}

Particle snapshots from HACC simulations contain particle positions, velocities, unique IDs, and information about the local gravitational potential. We saved three particle data sets: 1) the complete output from a single rank at each time step, 2) downsampled particle information at the 101 redshifts listed in Equation~(\ref{redshifts}) and 3) five full particle snapshots at $z=\{2.02, 1.01, 0.58, 0.15, 0.00\}$. These redshifts have proven to be useful when generating synthetic galaxy catalogs for large-scale surveys based on halo occupation distribution models.

The single rank output is mostly used for monitoring the health of the simulation while it is evolving. The data is written out in a format that can be readily processed by standard visualization tools such as ParaView~\citep{paraview} or VisIt~\citep{visit}. This allows us to create visualizations easily while the simulation is progressing. The data is also used to generate structure evolution movies. Figure~\ref{fig:evolv} illustrates rendered snapshots of Farpoint data. 

For the downsampled particle data set we save 1\% of the full particle snapshot. We randomly select the particles per rank for the first output and ensure the same subsample is saved at each of the subsequent snapshots. The particle data serves multiple post-processing functions. For example, correlation function measurements can be obtained from this subset of particles, as shown in Section~\ref{sec:corr}. The downsampled particle data can also be used to provide estimates of tidal fields, which in turn, can guide the modeling of intrinsic alignment studies (see, e.g., \citealt{2015PhR...558....1T} for a review). 

Finally, the full particle outputs have large storage requirements of $\sim$65TB per snapshot, and, therefore, it is only feasible to keep a handful. However, we measure summary statistics from the entire data set in situ, which are regularly output during the run. For example, we calculate the power spectrum at many redshifts. We provide a brief discussion and results for the power spectrum measurement in the following subsection.

\subsubsection{Matter Power Spectrum Measurement}
\label{sec:pow}

\begin{figure}[t]
\centerline{
 \includegraphics[width=3.5in]{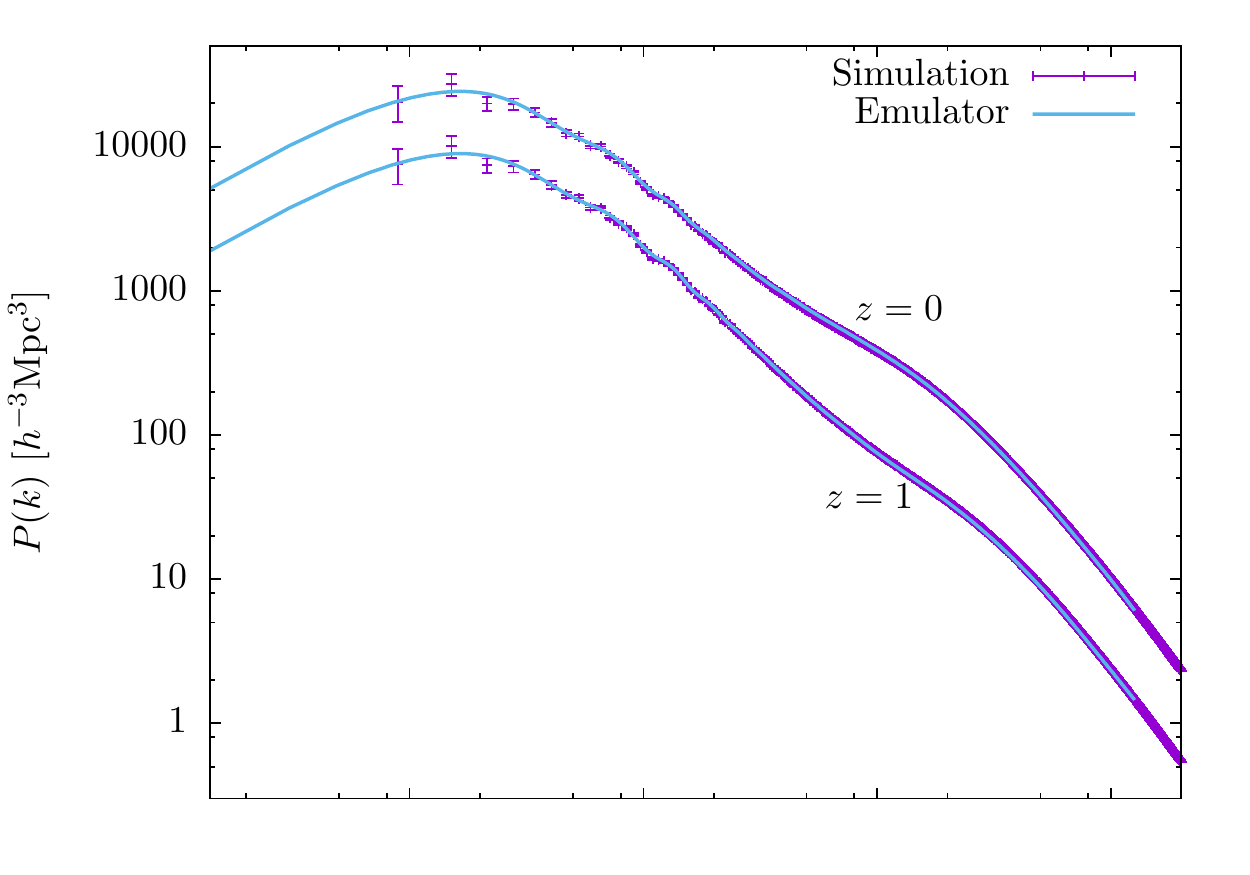}}
\vspace{-0.54cm}
\hspace{0.15cm}\centerline{\includegraphics[width=3.43in]{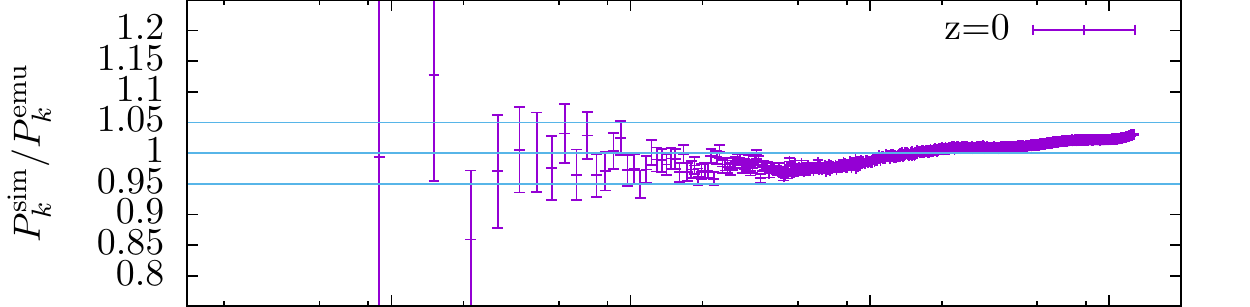}}
\vspace{-0.9cm}
\hspace{0.15cm}\centerline{\includegraphics[width=3.43in]{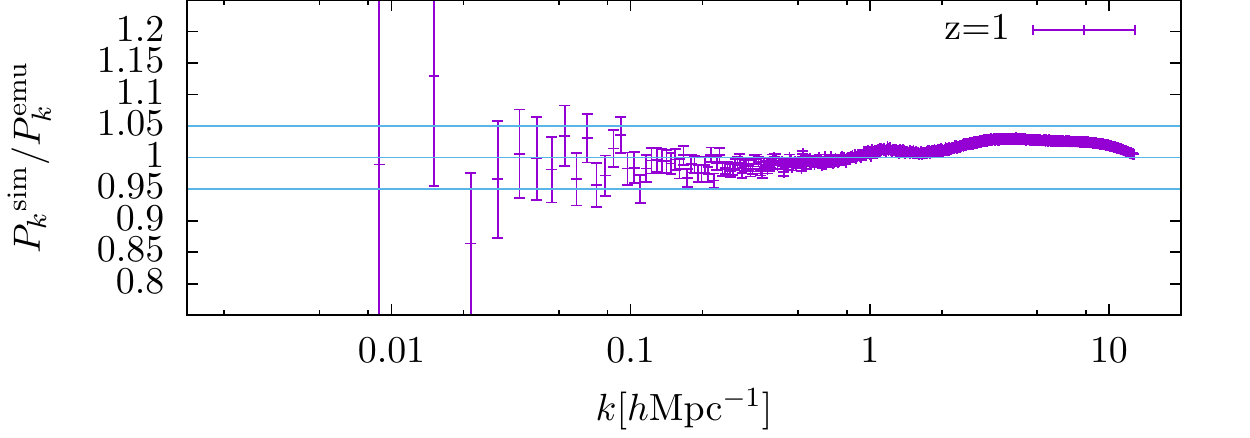}}
\vspace{0.5cm}
\caption{\label{fig:pow} Upper panel: Power spectrum measurements at redshift $z=0$ and
  $z=1$. Predictions from the Cosmic Emulator
  by~\cite{emu_ext} are shown for comparison. Lower panels: Ratio of the simulation and the emulator at both redshifts out to $k=10h$Mpc$^{-1}$, the maximum range for which the emulator was developed. The light blue bands indicate a 5\% range. The results are within the accuracy bounds reported in~\cite{emu_ext}.} 
\end{figure}

The nonlinear matter power spectrum provides an important measurement for extracting cosmological information. As widely discussed in the literature \citep{HutTak05,Hil08}, accurate predictions for the matter power spectrum and its evolution over time will be crucial to fully exploit the information delivered by next-generation surveys. In order to obtain high-accuracy predictions for the matter power spectrum, simulations have to fulfill stringent requirements with regard to volume and particle numbers. These requirements are discussed in, e.g.,~\cite{coyote1,Schneider16}. 

The resolution of the Farpoint simulation allows us to measure the power spectrum out to wavenumbers of $k\sim 10h$Mpc$^{-1}$ at high accuracy. The power spectrum is generated on the fly in HACC using a FFT-based method. The FFT grid size for the power spectrum was the same as for the particle-mesh (PM) solver, i.e., 12,228$^3$. Figure~\ref{fig:pow} shows the results for two redshifts, $z=0$ and $z=1$. In addition, we include the predictions from the Cosmic Emu emulator~\citep{emu_ext}. For the construction of the emulator, nested boxes were used in order to cover the full $k$-range of interest. The Farpoint simulation allows us to extract the same $k$-range from a single simulation. The overall agreement is better than 5\%, which is within the error bounds described in~\cite{emu_ext}. The estimate for the error bars of the power spectrum itself is described in~\cite{LJ1}.

\subsection{Particle Light Cones}
\label{sec:PLC}
\begin{figure*}[t]
\centerline{\includegraphics[width=6.5in]{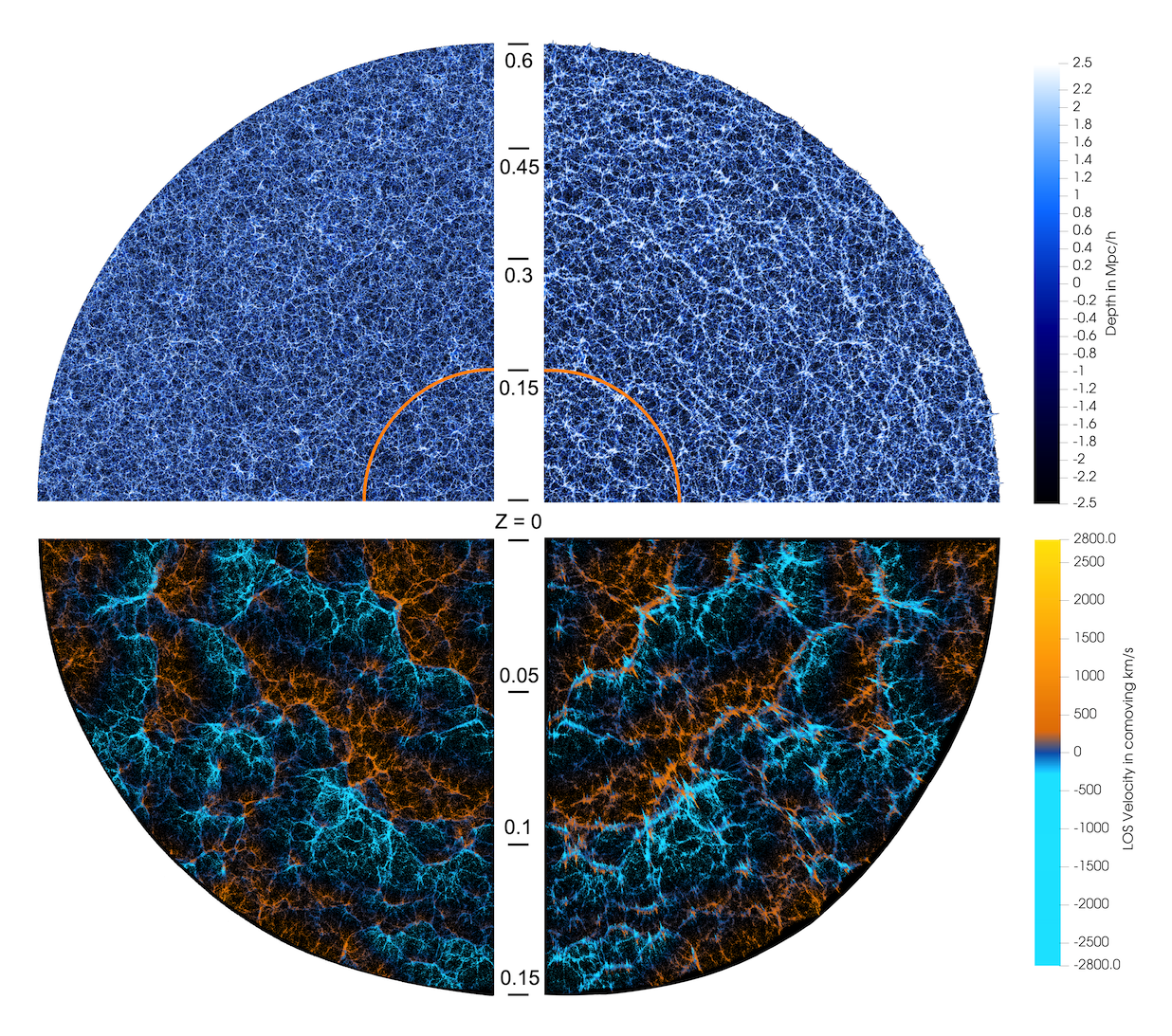}}
\caption{\label{fig:lc} Visualization of the particle light cone data out to $z=0.6$ (upper panels) and a zoom-in to $z=0.15$ (lower panels). The orange curve in the upper panels shows the $z=0.15$ boundary. The color in the upper panels represents depth in the disk of particles (in Mpc/h). In the lower panels, color represents LOS velocities (in comoving km/s). The left panels render the comoving particle positions while the right panels demonstrate the so-called \emph{Fingers of God} effect from redshift space distortion.}
\end{figure*}

In addition to storing particle snapshot data, we also generate particle light cones over a full sphere between redshifts $z=10$ and $z=0$. The observer is placed in the corner of the simulation box at coordinate $(0,0,0)$. Due to the finite volume of the simulation, replications of the box are necessary to cover the full light cone sphere at high redshifts. For accuracy, we interpolate a particle to the light cone frame by averaging a forward and backward extrapolation of the position at two encompassing times. We compute the average at each simulation time step to minimize interpolation error due to temporal discretization. Testing the light cone construction with different time step sizes, we have confirmed that the default simulation integrator incurs negligible discretization error on our data. 

Given the high mass resolution of the simulation, limited storage resources, and large particle replication at high redshifts, we downsampled the light cone particles to 0.1\% between redshifts $3 < z \leq 10$ and to 10\% for redshifts $z \leq 3$. 
For each particle we store the same attributes as for the snapshot data. The entire light cone output is 66TB for the high-redshift data and 2.9PB for the low-redshift range. The particle light cone data are further processed to generate density maps on a HEALPix\footnote{{\url{https://sourceforge.net/projects/healpix/}}}~\citep{healpix} grid of dimension Nside=16384. These outputs can be used for a range of analyses, such as building weak lensing maps, and reduce the total storage requirements of the data to 17TB (3TB and 14TB for the high and low redshift data, respectively). Generating the density maps in post-processing is very efficient and does not require access to the full machine. We confirmed that the selected downsampling of the particles leaves the pixels still adequately sampled. 

Figure~\ref{fig:lc} shows a visualization of the particle light cone data out to redshifts $z=0.6$ and $z=0.15$ in the upper and lower panels, respectively. The coloring in the upper panels shows the thickness of the visualized disk (in Mpc/h) while the colors in the lower panels represent line-of-sight (LOS) velocities (in comoving km/s). The left panels display the comoving positions of the particles, whereas the right panels include redshift space distortions; the elongation of structure oriented towards the observer in the right panels are a distortion effect often referred to as \emph{Fingers of God}. We describe our approach to generate the light cone image in more detail in the following.

To avoid rendering repeated structures in visualizing a thin disk out to $z\sim0.6$ (a comoving distance $d_c\sim1,500h^{-1}{\rm Mpc}$) from a periodic cubic simulation with side length of 1,000$h^{-1}$Mpc, we follow the volume remapping methodology of \cite{carlson}.
Through experimentation we determined that a disk thickness of 5$h^{-1}$Mpc provides a good visual density of structures. We used the {\it genremap} utility within the {\it BoxRemap}\footnote{\url{http://mwhite.berkeley.edu/BoxRemap/}} code to search for a suitable remapping, and we settled on the following set of integer lattice vectors
\begin{equation}
{\bf u}_1 = (7,7,3), {\bf u}_2 = (3,-6,7), {\bf u}_3 = (3,5,0),
\end{equation}
which corresponds to a volume-remapped cuboid with dimensions of 10,344$h^{-1}$Mpc by 9,695$h^{-1}$Mpc by 10$h^{-1}$Mpc. For this particular remapping the first two cuboid dimensions also have periodic boundary conditions, meaning the visualized disk could be extended indefinitely in radius with continuous structures, though features would start repeating after distances greater than $\sim10,000h^{-1}$Mpc. The third cuboid dimension does not have periodic boundary conditions, and we used a thickness of $\sim10h^{-1}$Mpc to provide some extra padding to decrease the likelihood of seeing two pieces of a split structure in different regions of the disk. 
We chose the normalized third cuboid directional vector as the normal vector to our visualization plane, $\hat{\bf n} = {\bf e}_3 / |{\bf e}_3|$ in the notation of \cite{carlson}. To select a thin disk of particles centered around this plane passing through the origin, we keep particles for which $| \hat{\bf n} \cdot {\bf r}| \leq 2.5h^{-1}{\rm Mpc}$, where ${\bf r}$ is the vector from the origin to the particle position in comoving distance units.

In order to visualize how the matter field would be distorted by observing in redshift space, we need to calculate the distance that would be measured from the observed redshift, which is a combination of the expansion of the universe since the photons were emitted and the LOS velocity. 
The original position of the particle can be expressed as ${\bf r} = d_c(z_{\rm emission}) \hat{\bf r}$, where $d_c$ is the comoving distance corresponding to a redshift $z_{\rm emission}$ (inferred via the Hubble distance-relation).
The LOS component of velocity is $v_{\rm LOS} = {\bf v} \cdot \hat{\bf r}$. HACC outputs particle peculiar velocities in comoving km/s. Since the speed of light is not a constant in comoving units, the redshift due to LOS velocity, $z_{\rm velocity}$, is computed as
\begin{eqnarray}
1 + z_{\rm velocity} & = & \sqrt{ \frac {1 + \frac {v_{\rm LOS}} {c(z_{\rm emission})}} {1 - \frac {v_{\rm LOS}} {c(z_{\rm emission})} } }, \\
c(z) & = & (1+z) c,
\end{eqnarray}
where $c$ is the usual constant value for the speed of light. The observed redshift, $z_{\rm observed}$, can then be calculated from the composition
\begin{equation}
1 + z_{\rm observed} = (1 + z_{\rm emission}) (1 + z_{\rm velocity}),
\end{equation}
which is used to calculate the redshift-space distorted position of the particle given by 
\begin{equation}
{\bf r}' = d_c(z_{\rm observed}) \hat{\bf r}.
\end{equation}
\\
\subsection{Halos and Cores}
\label{sec:haloscores}

Halo finding and characterization is a task well-suited to be carried out in situ. This processing does not significantly hamper the overall progress of the time stepper, and can leverage the full computational resource that is already being employed to evolve the simulation. Running the analysis on a smaller partition in post-processing would be inefficient, in addition to the writing, storing, and I/O costs being impractical in the case of an extreme-scale simulation. As part of the HACC framework, an OpenMP threaded halo finder is integrated into CosmoTools, and was run at the redshifts specified in Equation~(\ref{redshifts}) for the Farpoint campaign.

The halo finder ran a friends-of-friends (FOF) algorithm~\citep{davis85} using a linking length of $b=0.168$ to identify objects down to 50 particles per halo, or a halo mass of $M_{\rm FOF}=2.3\cdot 10^9 h^{-1}$M$_\odot$. The linking length was chosen to facilitate the generation of synthetic sky catalogs, similar to those discussed in
\cite{reid11} for redshift space distortion investigations, in \cite{white11} for measurements of the clustering of massive galaxies in the BOSS survey, and more recently for eBOSS investigations~\citep{eboss1,eboss2,eboss3,eboss4,eboss5,eboss6,eboss7}. For each FOF halo we store a range of properties, including a halo ID, mass, position (local gravitational potential minimum and center of mass) and velocity, kinetic energy, angular momentum, circular velocity and velocity dispersion, and the eigenvectors of the simple and reduced inertia tensor. In addition, we store the IDs for all the particles that belong to each halo, in order to enable the construction of halo merger trees in post-processing.

For each FOF halo above a size of 80 particles, we store the 50 particles closest to the halo center, i.e. the halo `core.' We mark these particles the first time they are found and track their evolution throughout the remainder of the simulation. Accordingly, we generate a file at each analysis time step, which contains the current core particles of each halo and their halo ID at that redshift, in addition to a separate file that accumulates and updates all core particles that have previously been identified. As we describe in Section~\ref{sec:core-merger}, we use this information to build detailed core merger trees that are capable of tracking substructure over time.

Finally, we also measure spherical overdensity (SO) halo properties. For each FOF halo with more than 500 particles, we measure $M_{200c}$ by growing spheres around the FOF gravitational potential center until the density falls below 200 times the critical density of the universe ($\rho_{\rm{c}}=3H^2/8\pi G$). This approach allows for efficient identification of SO halos and leads to accurate halo mass function measurements as shown in Section~\ref{sec:hmf}. For the SO halos, we store the same corresponding properties as was done for their FOF counterparts, in addition to providing concentration measurements obtained in three different ways~-- profile fitting, accumulated mass, and peak measurements. These methods are discussed in \cite{child} and we show the results for the profile fitting approach in Section~\ref{sec:cM}. \\

\subsubsection{Halo Mass Function Measurement}
\label{sec:hmf}

\begin{figure}[b]
\centerline{
 \includegraphics[width=3.2in]{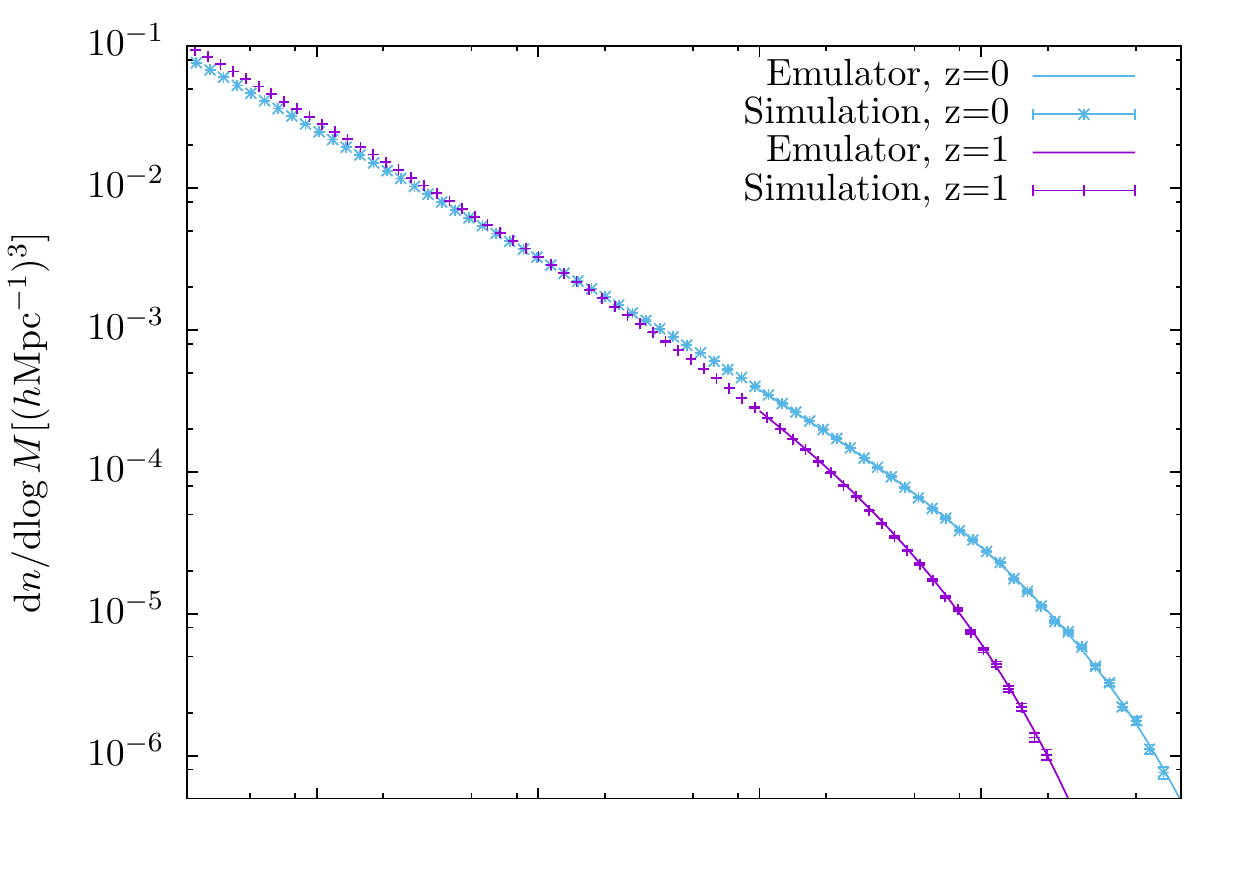}}
\vspace{-0.5cm}
\hspace{0.05cm}\centerline{\includegraphics[width=3.2in]{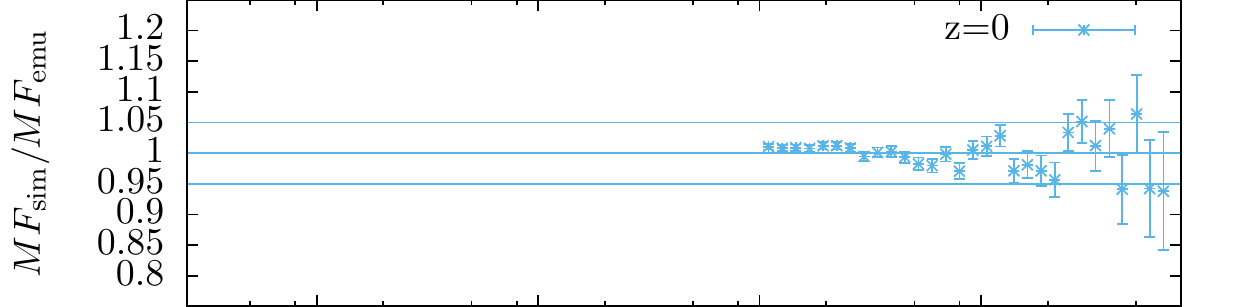}}
\vspace{-.9cm}
\hspace{0.05cm}\centerline{\includegraphics[width=3.2in]{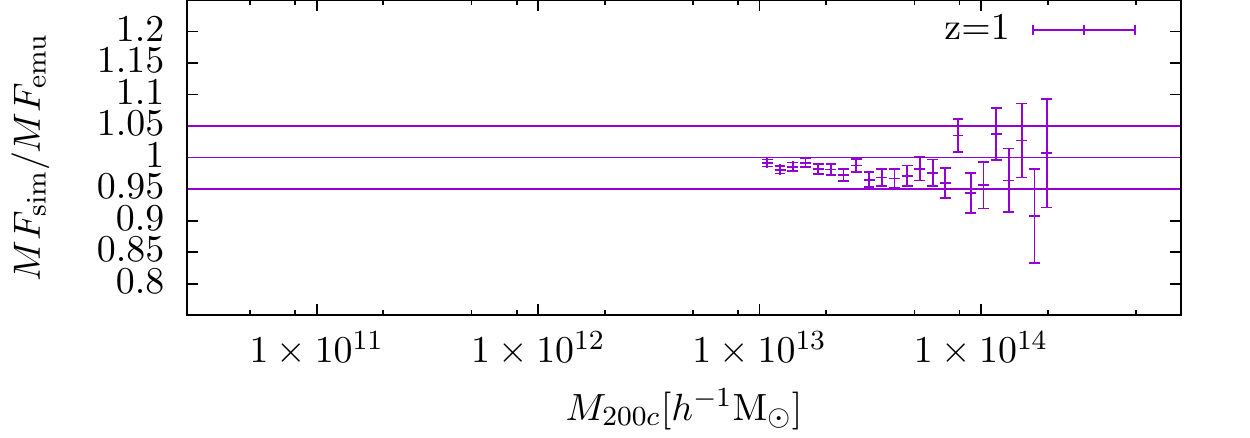}}
\vspace{0.5cm}
\caption{\label{fig:m200} Upper panel: Farpoint halo mass function measurements at redshift $z=0$ and $z=1$. We compare the results to the mass function emulator of ~\cite{emu_massf}.  Middle and Lower panel: Ratio of the simulation and the emulator results at both redshifts. Note that the emulator only provides predictions down to $M_{200c}\sim 10^{13} h^{-1}$M$_\odot$. For the mass range covered, the agreement is excellent and within the accuracy bounds reported in~\cite{emu_massf}.} 
\end{figure}

We discuss the results for the SO halo mass function at two redshifts, $z=1$ and $z=0$. The overdensity is measured with respect to 200$\rho_c$. The high mass resolution of the simulation allows us to identify halos down to a mass of $M_{200c}\sim 2.3\cdot 10^{10} h^{-1}$M$_\odot$, equivalent to 500 particles per halo. Figure~\ref{fig:m200} shows the mass function at the two redshifts. We ensure that each bin has at least 100 halos. For comparison, we include results from the mass function emulator based on the Mira-Titan simulation suite constructed by~\cite{emu_massf}. The set-up of the Mira-Titan Universe simulation suite is discussed in detail in~\cite{MT_1} and first results for power spectra predictions were reported in~\cite{MT_pk}. The simulations that were used to build the emulator had much lower mass resolution, resulting in a mass cut-off at $M_{200c}\sim 10^{13} h^{-1}$M$_\odot$. In the mass range covered by the emulator, we find close agreement with the Farpoint simulation.

\subsubsection{Concentration-Mass Relation}
\label{sec:cM}

\begin{figure}[b]
\hspace{0.13cm}\centerline{\includegraphics[width=3.42in]{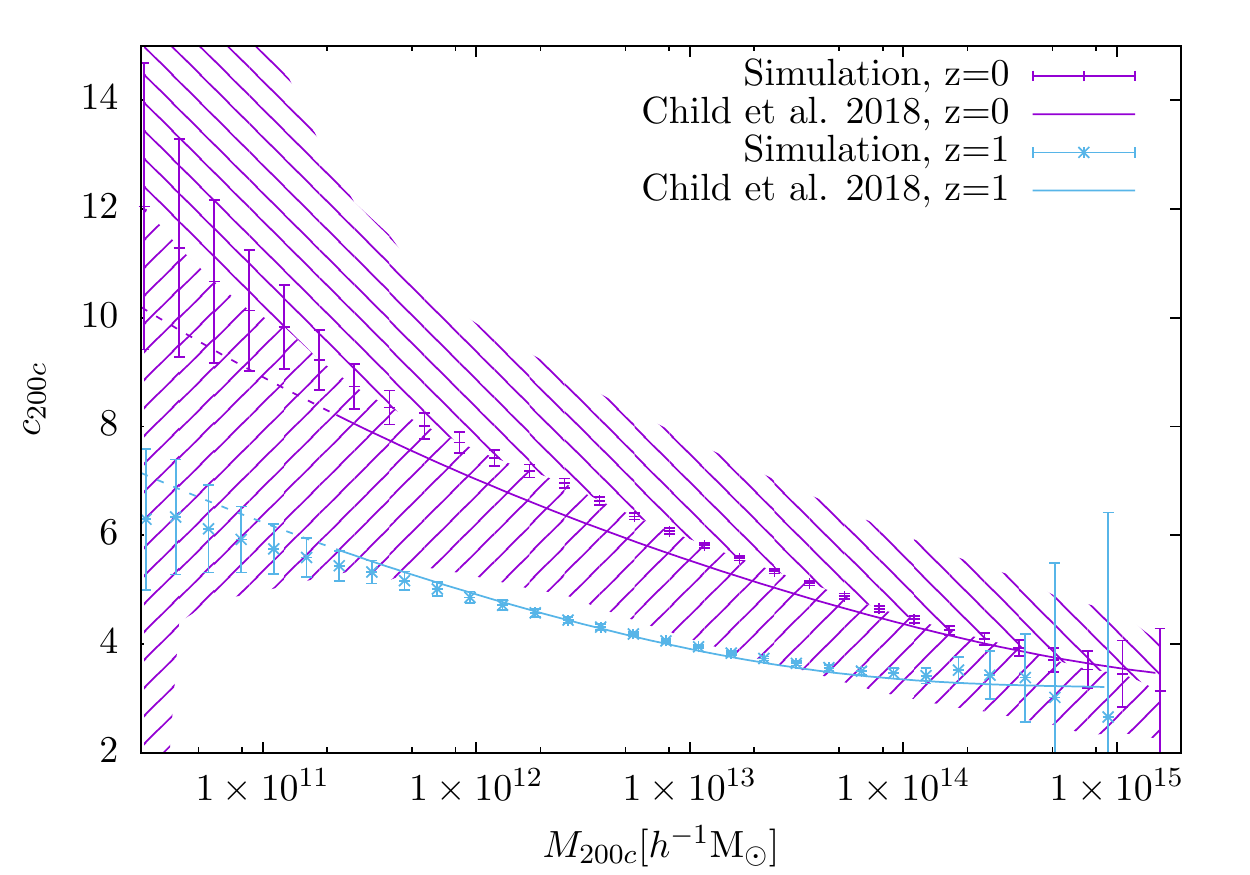}}
\caption{\label{fig:cm} Measurement of the halo concentration-mass relation at redshift $z=0$ (purple) and $z=1$ (light blue). The shaded region shows the 1-$\sigma$ standard deviation for $z=0$. For comparison, we show the fitting function derived in \cite{child} as implemented in the {\sc Colossus} package~\citep{diemer}. For the very low mass regime, the fit is represented by a dashed line, indicating that the results from \cite{child} were extrapolated in this mass range.} 
\end{figure}

We now present the results from the concentration-mass measurements at $z=0$ and $z=1$. The concentration-mass relation is derived from measurements of the SO halo profiles. We assume that every halo, relaxed or unrelaxed, is well-described by the Navarro-Frenk-White (NFW)-profile \citep{nfw1,nfw2}, given by:
\begin{equation}
\rho(r)= \frac{\delta\rho_{\rm{c}}}{(r/r_s)(1+r/r_s)^2}.
\label{eq:nfw}
\end{equation}
The NFW profile is described by $\delta$, the characteristic dimensionless density, and $r_s$, the scale radius. The concentration of a halo is then defined as the ratio $c_{\Delta}=r_{\Delta}/r_s$, where $r_{\Delta}$ is the radius at which the enclosed mass, $M_{\Delta}$, equals the volume of the sphere times $\Delta \rho_{\rm{c}}$. Here, $\Delta$ is the overdensity with respect to the critical density of the Universe (in our case, $\Delta=200$).

The concentration-mass relation has been extensively discussed in the literature, both in the context of observations and simulations. A range of fitting functions has been derived as well. These have been used to build models for the galaxy-halo connection and for comparison to observational data (see, e.g., \citealt{child} for a comprehensive compilation of observational results). The Farpoint simulation offers new measurements for low-mass halos due to its high mass resolution. At the same time, the gigaparsec volume of the simulation provides large halo counts across a wide halo mass range.

Figure~\ref{fig:cm} illustrates our measurements for two redshifts, $z=0$ and $z=1$. The Farpoint mass resolution permits measurements down to very small masses of $M_{200c}\sim2.7\cdot 10^{10} h^{-1}$M$_\odot$. The error bars shown for each bin are obtained by adding the error contribution from the individual concentration measurements and the Poisson error due to the finite number of halos in an individual bin in quadrature. A more detailed discussion can be found in \cite{bhattacharya13}. The two contributions to the error definition capture the uncertainty of the concentration measurement, in particular for low-mass halos, as well as the uncertainty due to low halo counts at large masses.

In addition to the measurements themselves, we also include the approximate fitting function derived in~\cite{child} (Equation~(18) and Table 1 in that paper). The fit is given as a function of $M_\star(z)$ to approximately capture the redshift and cosmology dependence of the concentration-mass relation. $M_\star(z)$ is the nonlinear ``collapse'' mass scale corresponding to peaks of the initial Gaussian random field collapsing at redshift z (Section 4 of~\cite{child} provides the definition details). We use the publicly available {\sc Colossus} package~\citep{diemer}, which includes a convenient conversion to the cosmology used in our simulation. The fit implemented in the package also allows for the extrapolation to smaller masses that were (deliberately) not covered in the original fit by~\cite{child}. However, we note the warning in~\cite{child} to not naively extrapolate the fit to masses smaller than those  originally considered. Given the large uncertainty due to the extrapolation, we have indicated this part of the fit via dashed lines.

As stated in~\cite{child}, the fitting function is not fully universal, but still provides a good estimate; our measurements in Figure~\ref{fig:cm} show that the fit is reasonably consistent with the Farpoint data, in particular for the results at $z=1$. 

\section{Post-processing Analysis}
\label{sec:post-analysis}

Not all types of analysis can be performed in situ, leaving the remaining tasks to be carried out in post-processing.
There are a number of reasons for this restriction. Some of our data products require the full history of the simulation to be available, such as the construction of merger trees. Other analysis codes are computationally expensive, yet do not necessarily demand the large processor allocation used in the simulation, e.g. measuring the matter correlation function. In addition, the post-processing investigations described in this section were carried out on reduced data sets, such as halo catalogs or downsampled particles, and, therefore, do not require the storage of the entire raw particle data.

\subsection{Correlation Function Measurement}
\label{sec:corr}

In this section we investigate the particle two-point correlation function~$\xi(r)$. Our approach for measuring the correlation function, as well as the error bars, is described in~\cite{LJ1}. We stored 1\% of the particle outputs at 101 redshifts as described in Section~\ref{sec:part}, leading to $\sim$19 billion particles per snapshot.

On large scales, even this reduced number of particles poses a major computational challenge for a correlation function measurement. However, the variance of the measured correlation function at these scales is dominated by the available simulation volume, rather than tracer density, allowing us to pursue a strategy of further subsampling the particles. We employed the same approximations as in~\cite{LJ1} to estimate the correlation function error bars on large scales; we use Gaussian terms with a non-linear power spectrum and a shot noise term for particle density -- a reasonable approximation on separations $\gtrsim 10 h^{-1}$Mpc. We estimate that on large scales ($>10 h^{-1}$Mpc) the total error bars are increased by less than $\sim$1\% when using only 0.01\% of the original simulation particle density.

Accordingly, we downsampled the output snapshot data~set by an additional factor of 100 and measured the correlation function over a distance range between $0.1 h^{-1}$Mpc to $125 h^{-1}$Mpc. To enhance performance, we implemented the pair counting approach on GPUs. The evaluation of $\xi(r)$ for $\sim$190M particles took $\sim$1000 sec on 8 nodes of Summit, which deploys 6 NVIDIA Volta GPUs per node. 
For small scales ($r\le 10 h^{-1}$Mpc), where we expect our estimated error bars to be less accurate, we also measured the correlation function from 0.5\% of the full particle set to test the fidelity of our results.
We found the difference between the 0.5\% and 0.01\% subsamples on small scales to be negligible. 

\begin{figure}[t]
\hspace{0.17cm}\vspace{-0.84cm}\centerline{\includegraphics[width=3.34in]{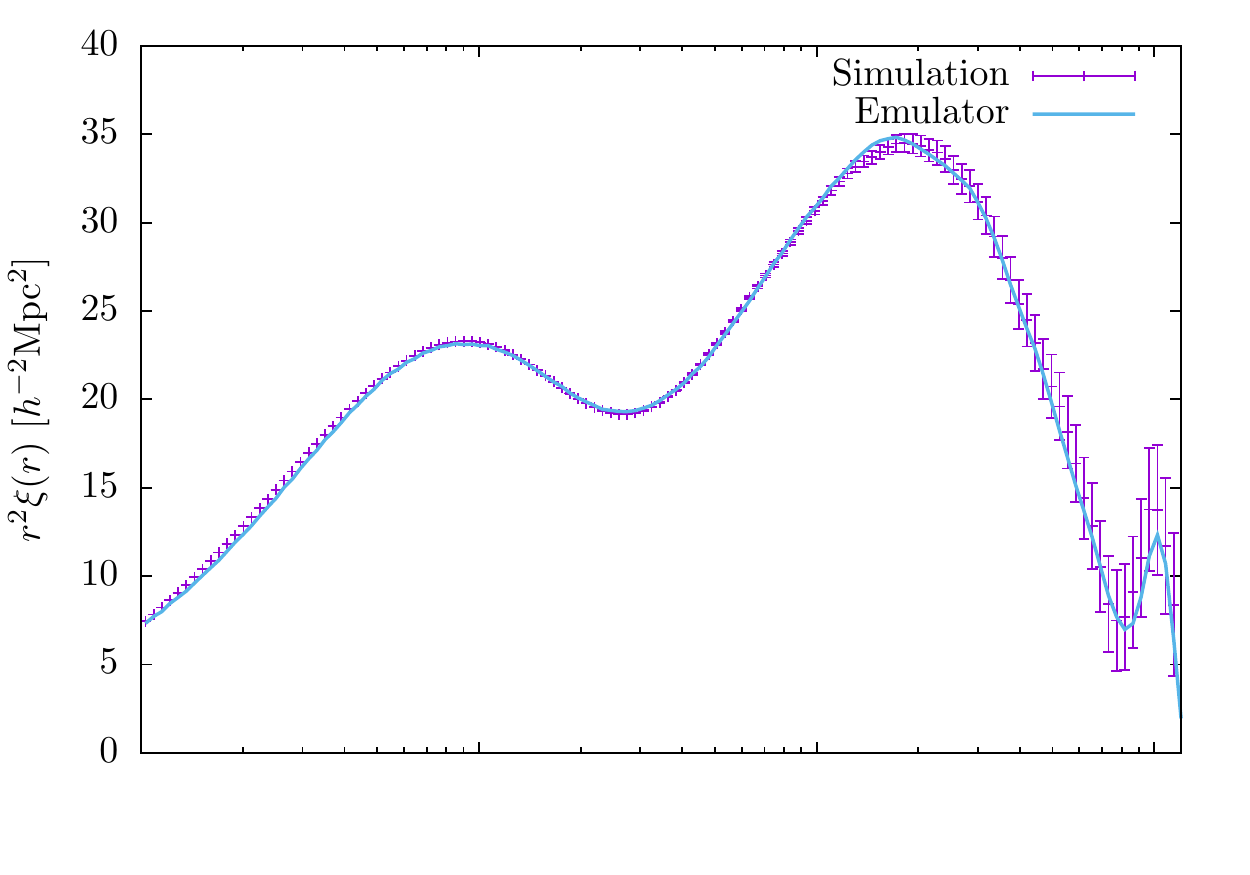}}
\centerline{\includegraphics[width=3.5in]{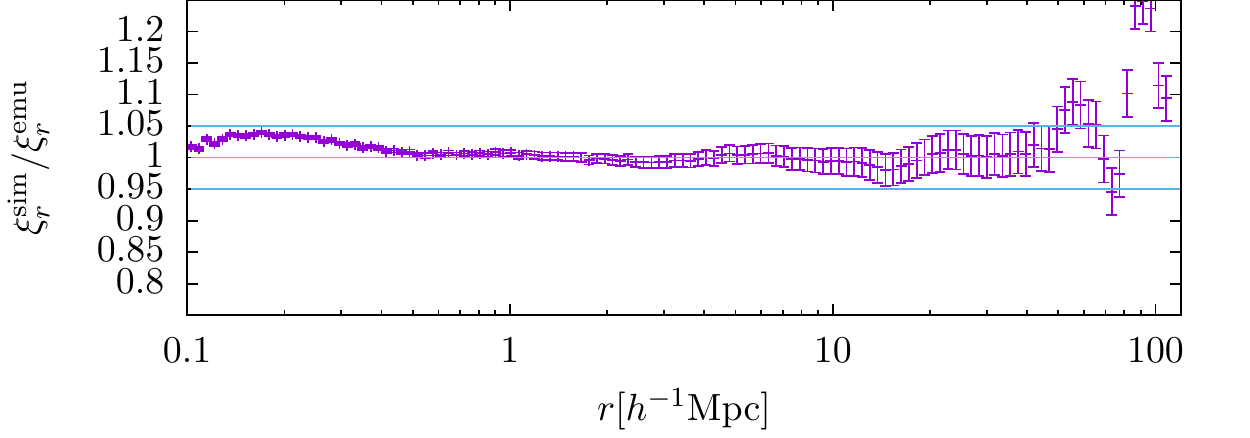}}
\caption{\label{fig:xi} The Farpoint particle correlation function at $z=0$. The measurement from the simulation is shown in purple and predictions from the Mira-Titan emulator described in~\cite{MT_pk} is shown in light blue. The BAO peak is visible at the right edge of the image. The ample resolution of the simulation allows us to evaluate the correlation function to scales down to 0.1$h^{-1}$Mpc.} 
\end{figure}

Figure~\ref{fig:xi} shows the measurement for $\xi(r)$ at $z=0$, in addition to the emulator predictions based on the power spectrum presented in~\cite{MT_pk} from the Mira-Titan Universe introduced in~\cite{MT_1}. We note that the result spans a wide range in $r$, from the BAO peak at $\sim$100$h^{-1}$Mpc down to small scales of $r\sim 0.1h^{-1}$Mpc. The agreement with the emulator is better than 5\% for most of the range considered, and is within the error estimates described in~\cite{MT_pk}. 

\subsection{Merger Trees}
\label{sec:mergertree}
An important data product generated in post-processing is the collection of halo merger trees.
Merger trees allow us to track the origin and evolution of halos and their properties over time, in addition to providing information about structure formation processes. They are also essential ingredients to semi-analytic models for creating synthetic sky catalogs. Driven by algorithmic considerations, our merger trees are built after the simulation is completed, where the evolution of the final halos that have formed are traced backwards in time. 

As discussed in Section~\ref{sec:haloscores}, we have identified halos using an FOF finder at 101 snapshots. For each halo, we have stored the IDs of the constituent particles. 
We follow the method described in~\cite{rangel16} to construct halo merger trees from this output; we match halo particles via their IDs across time, connecting overlapping halos of adjacent catalog snapshots.
The finite mass and time resolution of the simulation leads to some challenges that are discussed in detail in~\cite{rangel16} and in~\cite{LJ1}. We provide a brief summary here for completeness. 

One challenge concerns the minimum halo mass threshold that has to be chosen. A low-mass halo can be found in one snapshot but not reappear in the next due to mass loss, wherein the halo does not cross the detection threshold. This problem has been discussed extensively in the literature (see, e.g., \citealt{fakhouri08,behroozi13,sublink,han18} and references therein). We overcome this challenge by using a soft mass detection threshold. Once a halo has been identified, it can continue to exist in successive steps despite the possibility that it temporarily drops below the detection mass. This avoids the occurrence of disconnected tree branches. 

Another challenge concerns the treatment of halos that ``split'' and appear to have multiple descendants in subsequent snapshots. This effect is attributed to overlinking of FOF halos, whereby multiple proximate halos are identified as a single object even though they are not dynamically merged. In a following step, this unphysical link can break, leading to multiple halo descendants. Fly-by events can cause similar problems. In order to handle such situations, we use the concept of halo fragmentation -- we artificially break up halos that are later identified to have split. This allows us to create consistent merger trees without entangled branches. For more details, the reader is referred to~\cite{rangel16}. As described in the next section, halo merger trees are an important ingredient -- along with halo cores -- in our approach to halo substructure tracking.

\subsection{Core Catalogs}
\label{sec:core-merger}

Beyond the halo merger tree information, the high-resolution of the Farpoint run enables tracking of halo substructure over a wide dynamic range. Traditionally, substructure measurements employ subhalo finding approaches (for a comprehensive subhalo finder comparison project, see \citealt{2012MNRAS.423.1200O}). However, these methods can be computationally expensive, in addition to the subhalo definition being mutable. Thus, we follow a different strategy of identifying the center particles of a halo, rather than explicitly finding subhalos.
As explained in Section~\ref{sec:haloscores}, these so-called core particles are tracked throughout the evolution of the simulation. The combination of the halo merger trees, halo property files, and the core particle information is then used to build core merger trees, termed core catalogs.

The concept of core catalogs is described rigorously in~\cite{LJ2}, in which it is illustrated how cores can be used for substructure tracking. 
The assumption is that every substructure within a halo was once a distinct halo. During a merger event, the core particles of the captured halo can be marked as substructure in the new host halo. 
Several important properties are recorded on each core. 

First, the infall mass of the halo that a core originated from is stored; this information is important for estimating the (post-merger) mass evolution of substructure. Second, the core particle set position and velocity standard deviations are saved; these measurements can help determine if a substructure is still bound within a halo or has dispersed, as well as used to identify mergers between different substructures.

In a forthcoming paper (Korytov et al. in preparation) a careful study will be presented to model the galaxy distribution in clusters using cores. In that work, it is shown how the core mergers and disruption events within cluster environments can be modeled with the information supplied by core catalogs. The modeled galaxy distribution in clusters closely resembles observations from optical surveys.

We have created a core catalog for the Farpoint simulation. The data can be used for a range of projects that require substructure tracking, such as comprehensive modeling of the galaxy-halo connection, or structure formation studies. In the next subsection, we use the Farpoint core catalog to construct core mass functions, one of the many substructure studies achievable with this data product.

\subsubsection{Core Mass Function}

\begin{table*}[t] 
\begin{center}
\caption{Farpoint host halo mass bins}
\begin{tabular}{c|c|c|c}
$\log \left[ \langle M \rangle / \left(h^{{-1}}\mathrm{M_\odot} \right) \right]$ & $M_{min}$ ($10^{12}h^{{-1}}\mathrm{M_\odot}$)    & $M_{max}$ ($10^{12}h^{{-1}}\mathrm{M_\odot}$)   & Host halo count \\\hline\hline
12.5 (12.5) & 3.162 (3.162)     & 3.188 (3.191)     & 10001 (8677)          \\
13.0 (13.0) & 10.000 (10.000)   & 10.234 (10.355)   & 10000 (8770)           \\
13.5 (13.5) & 31.623 (31.625)   & 34.078 (36.364)   & 10000 (6369)           \\
14.1 (14.1) & 100.003 (100.060) & 139.115 (137.879) & 10000 (1395)           \\
14.5 (14.5) & 316.239 (316.478) & 374.719 (389.439) & 1000 (30)          
\end{tabular}\label{tab:hostbinsFP}
\end{center}
\begin{tablenotes}
    \item \small Halo masses and counts for the Farpoint simulation. The values given in each column indicate the $z=0$ host halo bins and are followed by numbers in parentheses for bins at $z=1$. Measurements of the core mass function (using the SMACC approach on the Farpoint data) for the different bins are shown in Figure~\ref{fig:cmf}. Note, we use very narrow mass bins due to the considerable statistics provided by the Farpoint simulation.
\end{tablenotes}
\end{table*}
In~\cite{LJ2} we introduced SMACC, the Subhalo Mass-loss Analysis using Core Catalogs approach to measure core mass functions from our generated catalogs. As described in the previous section, halo core tracking provides a fast and reliable method to follow substructure throughout the evolution of the simulation. In order to fully utilize the approach in, for example, semi-analytic models, we must determine the evolution of the substructure (core) masses after they have fallen into a host halo. Once calculated, the core masses can be added to the core catalogs to provide information similar to what is encapsulated in subhalo merger trees. The SMACC approach follows a model initially presented in~\cite{vdB2005}, which assumes that the average subhalo mass-loss rate is well-described by a power law: using the notation of~\cite{JvdB2016a},
\begin{equation} \label{eq:vdB_masslossrate}
    \dot{m}=-\mathcal{A}\frac{m}{\tau_{\mathrm{dyn}}}\left(\frac{m}{M}\right)^\zeta,
\end{equation}
where $M$ and $m$ are the parent halo mass and subhalo mass, respectively, $\tau_{\mathrm{dyn}}$ is the dynamical time of the halo, and $\{\mathcal{A},\zeta\}$ are free parameters. In~\cite{LJ2} we specify the optimization of $\{\mathcal{A},\zeta\}$ using results from an explicit subhalo finding approach.

\begin{figure}[t]
\centerline{\includegraphics[width=3.5in]{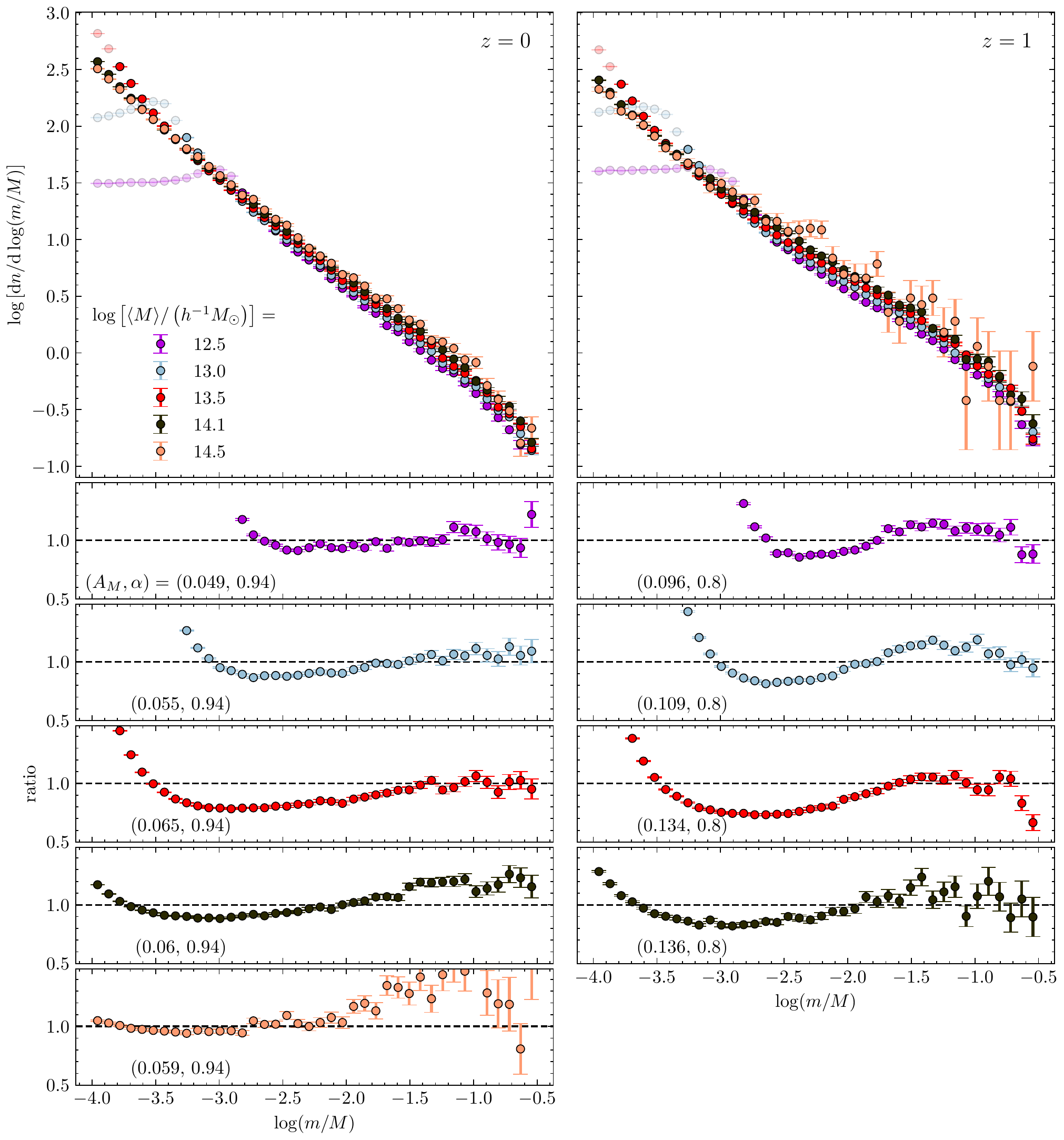}}
\caption{\label{fig:cmf}Results from the application of the SMACC approach to the Farpoint simulation.
Upper panels: Average core mass functions at $z=0$ (left) and $z=1$ (right) for five host halo mass bins. (Details of the host mass bins are listed in Table~\ref{tab:hostbinsFP}.) Points for which $\log(m/M)\le\log(100 m_{p,\mathrm{FP}}/\langle M \rangle)$ have enhanced color transparency, where $m_{p,\mathrm{FP}}$ is the particle mass of the Farpoint simulation and $\langle M \rangle$ is the average host halo mass of each host halo bin. Following \cite{LJ2}, for each host halo mass bin we fit Equation~(\ref{eq:fittingfn}) to the opaque data points with $\alpha$ fixed to $0.94$ and $0.8$ for $z=0$ and $z=1$, and allow $A_M$ to vary.
Lower panels: Ratio of the opaque points of the core mass function (shown in the upper panel) to the fitting function, for each host halo mass bin and redshift. The best fit parameter value for $A_M$ is indicated in each panel.}
\end{figure}

We applied SMACC to the Farpoint simulation, using the fiducial parameters $(\mathcal{A},\zeta)=(1.1, 0.1)$ determined in~\cite{LJ2}. 
The resulting core mass functions for $z=0$ and $z=1$ are shown in Figure~\ref{fig:cmf}. We present measurements for the five host halo mass bins listed in Table~\ref{tab:hostbinsFP}. Due to the substantial statistics provided by the Farpoint simulation, we are able to choose very narrow mass bins. Overall, the results are consistent with the core mass function measurements shown in~\cite{LJ2}.

We fit Equation~(6) of~\cite{vdB16} to our core mass functions:
\begin{equation}\label{eq:fittingfn}
\frac{\mathrm{d}N}{\mathrm{d}\log(m/M)}=A_M\left(\frac{m}{M}\right)^{-\alpha}\exp\left[-50(m/M)^4\right].
\end{equation}
For consistency with \cite{LJ2}, we fixed $\alpha=(0.94,0.8)$ for $z=(0,1)$ and allowed $A_M$ to vary.
The lower panels of Figure~\ref{fig:cmf} show ratios of the core mass functions to the fitting functions for each host halo mass bin; the best fit values of $A_M$ are indicated in each panel. Note that we omitted the highest host halo mass bin at $z=1$ from the fitting procedure due to the low host halo count. As was found for the Last Journey simulation in \cite{LJ2}, Equation~(\ref{eq:fittingfn}) provides a reasonable fit to our data.  

\subsection{Halo Light Cones}

An additional data product generated in post-processing is the halo light cone. Halo light cones are essential inputs to the generation of synthetic sky simulations, allowing for the creation of maps from the viewpoint of an observer. 
In order to construct halo light cones, we combine information from the halo merger trees and the halo catalogs at the 101 snapshots listed in Equation~(\ref{redshifts}). 

The merger trees allow us to identify halos and their progenitor's position at the previous snapshot, and use this information to interpolate positions backward onto the light cone. If a halo has multiple progenitors, we choose the most massive for the interpolation. If a halo has no progenitor -- this could happen if, e.g., the halo mass fell below the detection threshold -- we extrapolate the position backward onto the light cone using the velocity at the later snapshot. Fragment halos (mentioned at the end of Section~\ref{sec:mergertree}) are treated separately; for each set of halo fragments, we retain the most massive fragment and discard all other objects associated with the halo, and then use the merger tree information to interpolate or extrapolate as described above. We then assign properties to the fragment object by matching against the halo catalog at the later snapshot.

Similar to the particle light cones described in Section ~\ref{sec:PLC}, we generate a full sphere for the halo light cones, placing an observer in the corner of the simulation box at coordinate $(0,0,0)$. The finite volume of the Farpoint simulation requires duplications of the box at higher redshifts.

\section{Data Release}
\label{sec:data}

We introduced the HACC Simulation Data Portal\footnote{\url{https://cosmology.alcf.anl.gov/}} in \cite{heitmann19a}. The portal provides a web-based interface that allows for easy access to a subset of the simulation outputs. It is set up on  Petrel\footnote{\url{https://press3.mcs.anl.gov/petrel}}, a pilot infrastructure for data  management and sharing hosted at the Argonne Leadership Computing Facility (ALCF). Data sets can be selected via a drop-down menu for different redshifts, triggering data transfers using the Globus service\footnote{\url{https://www.globus.org/}}. The equivalent data products that we released as part of the Last Journey simulation~\citep{LJ1} are available for the Farpoint run. Specifically, we release outputs for nine redshifts: 
\begin{equation}
z=\{0.0,0.05,0.21,0.50, 0.54, 0.78, 0.86, 1.43, 1.49\}.
\end{equation} 
For each redshift, halo properties measured with an FOF halo finder using a linking length of $b=0.168$ are available as follows: 
halo ID, $M_{\rm halo}$, $(x,y,z)_{\rm pot}$, $(x,y,z)_{\rm COM}$, and $(v_x,v_y,v_z)_{\rm COM}$. Halo masses are measured in $h^{-1}$M$_\odot$, position are given in comoving $h^{-1}$Mpc, and velocities in comoving peculiar~km/s. We note that the halo IDs are not kept the same between redshifts. Accordingly, the halos cannot be traced over time without additional information from the particle files. For the center definition, we provide both the gravitational potential minimum and the center of mass measurement; these two quantities can differ significantly for large, unrelaxed halos. 

We release information about the particles residing within individual halos. In order to keep the file sizes manageable, we stored 1\% of the constituent particles per halo; small halos with a total mass less than $500 m_p$ store 5 particles.  
For each particle, positions, velocities, and halo and particle IDs are recorded. The units are the same as for the halo properties. The halo IDs allow the connection of the halo particles to their host halos at each time step. The particle IDs are consistent across redshifts. However, the particles inside halos are randomly chosen for each snapshot separately, so there is no guarantee that the same particles can be identified across redshifts.

Lastly, the Portal provides access to 1\% of the raw particle data randomly sampled for each redshift, where the selection is identical between all snapshots. 

\section{Summary and Outlook}
\label{sec:summary}

In this paper we introduced the Farpoint simulation, a new member of the HACC extreme-scale simulation suite. The underlying cosmology parameters are the same as for the recently released Last Journey simulation~\citep{LJ1}, albeit at a much higher mass resolution of $m_p\sim 4.6\cdot 10^7 h^{-1}$M$_\odot$. We provided a range of basic measurements, including results for the matter power spectra, matter correlation function, halo mass function and the concentration-mass relation. We showed comparisons to commonly used emulators over the mass and length scales where predictions are available. Overall, we found very favorable agreement, consistent with the error estimates provided by the emulators. We also discussed the particle and halo light cone data products, in addition to the detailed halo catalogs.  

The high mass resolution of the simulation allows for comprehensive tracking of substructures throughout the structure formation process. To that end, we employed the halo core tracking strategy extensively discussed in \cite{rangel16} and \cite{LJ1}, generating core catalogs for the Farpoint run. From the catalogs, we measured the core mass function using the mass loss model discussed in~\cite{LJ2}. Finally, we made a subset of the Farpoint data products publicly available through the HACC Simulation Data Portal. 

The Farpoint campaign will facilitate a range of follow-on projects. In particular, studies of the galaxy-halo connection will be of major interest, given the mass resolution and available volume of the simulation. \cite{wechsler18} provide an overview of the topic and list future challenges. As a specific example, investigations on the galaxy-halo assembly correlation as carried out in \cite{2019MNRAS.488.3143B} are very interesting and can potentially benefit from better statistics and higher resolution. 
Models of the halo assembly history and assembly bias studies as discussed in, e.g.~\cite{hearin2021}, can also exploit the higher resolution of Farpoint, in addition to the improved statistics afforded by the factor of $\sim 60\times$ increase in volume. 

The availability of the Last Journey simulation in a much larger volume at lower mass resolution will enable important convergence studies with Farpoint as well. For example, in \cite{cosmoDC2}, very faint galaxies were distributed without a halo assignment, since low mass host halos were not resolved. By comparing the Farpoint and Last Journey simulations, this approach can be tested in detail. 

Another important area concerns the study of the evolution of halos and their properties over time. The excellent mass resolution of the Farpoint simulation provides information about halo formation and substructure evolution processes starting at early times. While focusing on static measurements in this paper, we have dynamic data products in hand -- such as merger trees and core catalogs -- that will enable such additional studies. These and other topics will be investigated in forthcoming papers.

\begin{acknowledgments}

Argonne National Laboratory's work was supported under the U.S. Department of Energy contract DE-AC02-06CH11357.  Awards of computer time were provided by the ASCR Leadership Computing Challenge (ALCC) program.  This research used resources of the Oak Ridge Leadership Computing Facility, which is a DOE Office of Science User Facility supported under Contract DE-AC05-00OR22725. We are indebted to the OLCF team for their outstanding support and help to enable us to carry out a simulation at this scale. This research was supported by the Exascale Computing Project (17-SC-20-SC), a collaborative effort of the U.S. Department of Energy Office of Science and the National Nuclear Security Administration. This research used resources of the Argonne Leadership Computing Facility, which is a DOE Office of Science User Facility supported under Contract DE-AC02-06CH11357.

\end{acknowledgments}

\software{
HACC~\citep{habib14}, 
Colossus~\citep{diemer}, 
BoxRemap~\citep{carlson} \url{http://mwhite.berkeley.edu/BoxRemap/}, 
HEALPix~\citep{healpix} \url{https://sourceforge.net/projects/healpix/}, 
vl3~\citep{rizzi2014performance}, 
ParaView~\citep{paraview}, 
Blosc library \url{https://blosc.org/}, 
HACC Simulation Data Portal \url{https://cosmology.alcf.anl.gov/}, 
Globus service \url{https://www.globus.org/}, 
2D-FFTLog~\citep{2dfftlog} \url{https://github.com/xfangcosmo/2DFFTLog}
}

\end{document}